\documentclass[10pt,amsmath,amssymb,pra,aps,twocolumn]{revtex4-2}

\usepackage[dvips]{graphicx}% Include figure files
\usepackage{dcolumn}% Align table columns on decimal point
\usepackage{bm}% bold math
\usepackage[colorlinks=true, allcolors=blue]{hyperref}
\usepackage{slashed}
\usepackage{multirow}
\usepackage{feynmf}
\usepackage{diagbox}
\usepackage{color,soul}
\usepackage{ulem}
\definecolor{BLUE}{rgb}{1.0,0.0,1.0}

\usepackage{tablefootnote}
\usepackage{natbib}
\begin{document}

\title{Exploring the Properties of Light Diatomic Molecules in Strong Magnetic Fields}% Force line breaks with \\
\author{ T. Zalialiutdinov$^{\,1,\,2}$}
\email[E-mail:]{t.zalialiutdinov@spbu.ru}
\author{D. Solovyev$^{\,1,\,2}$}
\affiliation{ 
$^1$ Department of Physics, St. Petersburg State University, Petrodvorets, Oulianovskaya 1, 198504, St. Petersburg, Russia \\
$^2$ Petersburg Nuclear Physics Institute named by B.P. Konstantinov of National Research Centre 'Kurchatov Institut', St. Petersburg, Gatchina 188300, Russia \\
}

\begin{abstract}
In this study, we develop and implement a specialized coupled-cluster (CC) approach tailored for accurately describing atoms and molecules in strong magnetic fields. Using the open-source Ghent Quantum Chemistry Package (\texttt{GQCP}) in conjunction with the Python-based Simulations of Chemistry Framework (\texttt{PySCF}), we calculate potential energy curves, permanent and transient dipole moments, as well as vibrational spectra for the diatomic molecules H$_2$, HeH$^+$ and LiH under various magnetic field strengths adopting a fully non-perturbative treatment. The main computational difficulties stem from the inclusion of the magnetic field in the Hamiltonian, in particular, from the presence of the angular momentum operator, which leads to a complication of the wave function and introduces a gauge-origin dependence. Addressing these challenges requires advanced modifications to existing routines, which we achieve by implementing gauge-including atomic orbitals (GIAOs) by using \texttt{GQCP}, and the capabilities offered by \texttt{PySCF}. This approach enhances the accuracy and reliability of the CC theory, opening pathways for more comprehensive investigations in molecular quantum chemistry at strong magnetic fields.
\end{abstract}

%\keywords{Suggested keywords}%Use showkeys class option if keyword
                              %display desired
\maketitle

%\tableofcontents

\section{Introduction}

The study of atoms and molecules subjected to strong magnetic fields has garnered substantial interest in recent years \cite{Turbiner2021, Pemberton2022, RevModPhys.73.629}. Although highly homogeneous magnetic fields of the order of $10^3$ Tesla (equivalent to $0.043$ in atomic units, where one atomic unit of field strength is $2.35 \times 10^5$ T) have not yet been generated in laboratories, such extreme conditions occur naturally in the atmospheres of magnetised white dwarf stars, for example. Observations of helium \cite{Jordan2001}, hydrogen molecules \cite{Bignami2003, Sanwal2002}, and more recently heavier elements \cite{Hollands2020, Hollands2023} in these astrophysical environments underscore the critical importance of exploring the behavior of atoms and molecules under intense magnetic fields \cite{Ferrario2020}. This line of research is essential for advancing our understanding of fundamental physical and chemical processes and also plays a pivotal role in astrophysics. In particular, it assists in the interpretation of the observational spectra of magnetic white dwarfs and neutron stars, thereby facilitating the determination of their magnetic field strengths \cite{Berdyugin2023, Rosato_2019}. 

Although the strongest sustained magnetic fields generated in terrestrial laboratories are on the order of $10$ Tesla, this value is generally small enough that such fields can typically be treated as a perturbation. This assumption breaks down as the field strength approaches $1$ a.u., where the underlying physics becomes more complex. In such a case, the magnetic forces become comparable in magnitude to the Coulomb forces, requiring a non-perturbative treatment \cite{Stopkowicz_2015, Tellgren_2014}. Furthermore, such conditions give rise to a new type of chemical bond known as a paramagnetic perpendicular bonding \cite{Lange2012}. 

The investigation of atoms and molecules in strong fields is accompanied with considerable difficulties, arising primarily from the limitations inherent in the existing theoretical methods and the complex nature of the quantum systems under study. The need for reliable theoretical models and accurate quantum-chemical calculations is particularly pronounced, owing to the complex behavior of electron correlation within these extreme conditions.

During the past decade, the Full Configuration Interaction (FCI) and Coupled Cluster (CC) methods have emerged as the prevalent approaches for such studies, as evidenced by several groundbreaking works in the field \cite{BAGEL, Stopkowicz_2015, Kitsaras_2024}. However, the computational demands of FCI are extraordinarily high, rendering it feasible only for very small systems with a limited number of electrons. This substantial computational cost is attributed to the requirement to consider all possible electron configurations, which increase exponentially with the size of the system.
Given these constraints, there is a pressing need to develop and refine quantum-chemical methods capable of handling larger and more complex systems with a similar level of accuracy. Advancements in this area would not only facilitate the study of larger molecules and systems with more electrons but could also lead to the discovery of new phenomena or provide deeper insights into the behavior of matter under strong magnetic fields. 

Recently, significant progress has been made in the realization of non-perturbative magnetic field effects within coupled-cluster (CC) methods \cite{Culpitt_2023, Tellgren_2014, Stopkowicz_2015, arxiv}. In particular, several specialized software packages have been developed, including \texttt{ChronusQ} (Chronus Quantum) \cite{ChronusQ}, \texttt{HelFEM} (Helsinki Finite Element Suite for Atoms and Diatomic Molecules) \cite{Lehtola2019}, \texttt{BAGEL} (Brilliantly Advanced General Electronic-structure Library) \cite{BAGEL}, \texttt{QCumbre} (Quantum Chemical Utility Enabling Magnetic-Field Dependent Investigations Benefiting from Rigorous Electron-Correlation Treatment) \cite{QCumbre}, and the \texttt{LONDON} code \cite{LONDON}.

Among these software packages, \texttt{QCumbre} and \texttt{LONDON} are notable for their ability to incorporate electron correlation effects through the application of the coupled-cluster approach. Additionally, \texttt{BAGEL} offers capabilities for relativistic calculations within the framework of the full configuration interaction (FCI) method. However, this latter capability comes at the cost of significant restrictions on the size of the systems that can be considered, owing to the substantial computational demands required.

In this study, we present our computational implementation for calculating binding energies and electronic structure properties of HeH$^+$, LiH, and H$_2$ molecules exposed to strong magnetic fields. The approach utilizes complex coupled-cluster methods and is built on the open-source libraries available through The Ghent Quantum Chemistry Package (GQCP) \cite{GQCP} and the Python-based Simulations of Chemistry Framework (PySCF) \cite{PySCF}. Throughout the paper, atomic units are used, where $\hbar=e=m_e=1$ ($\hbar$ is the Planck constant, $e$ is the charge of an electron and $m_e$ is the electron mass). 

\section{Theory}
\label{h1}

Within the Born-Oppenheimer approximation, the Hamiltonian of a molecule with $N$-electrons subjected to a uniform magnetic field can be represented as follows
\begin{eqnarray}
\label{00}
    \hat{H} = \hat{H}_0 + \sum_{i=1}^{N}\left( \mathbf{A}(\mathbf{r}_i) \cdot \mathbf{p}_i + 
    %\sum_{i=1}^{N}
    \mathbf{B}(\mathbf{r}_i) \cdot \mathbf{s}_i
%   \\\nonumber
    + \frac{1}{2} %\sum_{i=1}^{N}
    \mathbf{A}^2(\mathbf{r}_i)\right).\,\,\,
\end{eqnarray}
Here the usual field-free electronic Hamiltonian, denoted as $ \hat{H}_0 $, is
\begin{eqnarray}
    \label{H0}
    \hat{H}_0 = -\sum_{i=1}^N \frac{1}{2} \nabla_i^2 + \sum_{i=1}^N \sum_{j>i}^N \frac{1}{|\mathbf{r}_i - \mathbf{r}_j|} 
    \\
    \nonumber
    - \sum_{i=1}^N \sum_{A=1}^M \frac{Z_A}{|\mathbf{r}_i - \mathbf{R}_A|}
    ,
\end{eqnarray}
which includes the kinetic energy operator for the electrons and the potential energy contributions arising from electron-electron and electron-nucleus interactions ($N$ is the number of electrons, $M$ is the number of nuclei). Note that in Eq.~(\ref{00}), the Coulomb gauge, where $ \nabla \cdot \mathbf{A} = 0 $, is assumed.
The magnetic field is given by the vector potential $\mathbf{A}$ with the relation $\mathbf{B} = [\nabla \times \mathbf{A}] = \mathrm{const}$. The vector product is denoted by a times sign (as opposed to the scalar product of vectors, which is indicated by a dot), and vectors are given in bold. In the additional terms, the vector potential can be expressed as $ \mathbf{A}(\mathbf{r}_i) = \frac{1}{2} [\mathbf{B} \times (\mathbf{r}_i - \mathbf{O})] = \frac{1}{2} [\mathbf{B} \times \mathbf{r}^{\mathrm{O}}_i ]$, where $\mathbf{s}_{i}$ is the electron spin, $\mathbf{r}_i^{\mathrm{O}} = \mathbf{r}_{i} - \mathbf{O} $ is the position vector of the $i$-th electron relative to the global gauge origin $\mathbf{O}$ (the gauge origin is arbitrary) \cite{Stopkowicz_2015}.

The second term in Eq.~(\ref{00}), which is known as orbital paramagnetic interaction, can be transformed to 
\begin{eqnarray}
 \mathbf{A}(\mathbf{r}_i) \cdot \mathbf{p}_i = \frac{1}{2} [\mathbf{B} \times (\mathbf{r}_i - \mathbf{O})] \cdot \mathbf{p}_i 
 \\\nonumber
 = \frac{1}{2} \mathbf{B} \cdot \left[(\mathbf{r}_i - \mathbf{O}) \times \mathbf{p}_i\right] 
 = \frac{1}{2} \mathbf{B} \cdot \mathbf{l}_i^{\mathrm{O}},   
\end{eqnarray}
where $\mathbf{l}_i^{\mathrm{O}} = -\mathrm{i}[\mathbf{r}_i^{\mathrm{O}} \times \nabla_i]$ represents the angular momentum operator relative to the gauge origin. The third term in Eq.~(\ref{00}) corresponds to the spin paramagnetic interaction, and the fourth term (diamagnetic interaction) can be rewritten as follows:
\begin{eqnarray}
\frac{1}{2} \mathbf{A}^2(\mathbf{r}_i) = \frac{1}{8}
[\mathbf{B} \times \mathbf{r}^{\mathrm{O}}_{i}]\cdot [\mathbf{B} \times \mathbf{r}^{\mathrm{O}}_{i}]
\\\nonumber
= B^2 (r_i^{\mathrm{O}})^2 - (\mathbf{B} \cdot \mathbf{r}_i^{\mathrm{O}})^2.    
\end{eqnarray}
Thus, the gauge origin dependent Hamiltonian can be expressed as:
\begin{eqnarray}
\label{1}
    \hat{H} = \hat{H}_0 + \frac{1}{2} \sum_{i=1}^{N} \mathbf{B} \cdot \mathbf{l}_i^{\mathrm{O}} + \mathbf{B} \cdot \mathbf{S} 
    \\\nonumber
    + \frac{1}{8} \sum_{i=1}^{N} \left[ B^2 (r_i^{\mathrm{O}})^2 - (\mathbf{B} \cdot \mathbf{r}_i^{\mathrm{O}})^2 \right],
\end{eqnarray}
where $\mathbf{S}$ is the total spin of the system. 

The presence of the position vector $\mathbf{r}_i^{\mathrm{O}}$ in the Hamiltonian (\ref{1}) is directly relevant to the definition of the gauge-dependent vector potential. It is easy to see that different choices of gauge origin, $\bm{O}\rightarrow \bm{G}$, in the external vector potential $ \mathbf{A}(\mathbf{r}_i) = \frac{1}{2} [\mathbf{B} \times (\mathbf{r}_i - \mathbf{O})] $ are related by gauge transformation: 
\begin{eqnarray}
\label{G}
\bm{A}_{\bm{G}} (\bm{r}_i)= \frac{1}{2}\bm{B} \times (\bm{r}_i - \bm{G}) 
\\\nonumber
= \frac{1}{2}\bm{B} \times (\bm{r}_i - \bm{O}) - \frac{1}{2}\bm{B} \times (\bm{G} - \bm{O})   
\\\nonumber
=\bm{A}_{{\bm{O}}} (\bm{r}_i) - \bm{A}_{\bm O} (\bm{G})\equiv \bm{A}_{\bm O} (\bm{r}_i) + \nabla f, 
\end{eqnarray}
with $ f(\bm{r}_i)= -\bm{A}_{\bm O}(\bm{G}) \cdot \bm{r}_i  $. It is also known that the wave functions $\Psi$ of a particle in an electromagnetic field are not uniquely defined, because the choice of the field potentials is not unique: they are defined only to within a gauge transformation, see \cite{Landau:1991wop}.

The conversion Eq.~(\ref{G}) does not affect the values of the field strengths, and it is therefore evident that it cannot fundamentally alter the solutions of the wave equation. In particular, such a transformation should leave $|\Psi|^2$ unaltered, i.e., it should be manifested as a phase of the wave function. The invariance of the original equation is restored by simultaneously modifying the Hamiltonian according to Eq.~(\ref{G}) and adding the corresponding phase shift to the wave function.  The non-uniqueness of the wave function does not affect any physically significant quantities. This is because they do not explicitly depend on the potentials, since they are derived by averaging some operators. As a result, any complex conjugate phases cancel out, ensuring that the physical observables remain unaffected. Consequently, the exact wave function of an electron in a magnetic field according to Eq.~(\ref{G}) should be modified to
\begin{eqnarray}
    \label{psi}
    \Psi_\mathrm{G} = \exp{[\mathrm{i} \mathbf{A}(\mathbf{G})\cdot\mathbf{r} ]} \Psi_\mathrm{O}
    \\\nonumber
    =
    \exp{\left[\frac{\mathrm{i}}{2}\textbf{B}\times (\textbf{G}-\textbf{O}) \cdot\mathbf{r} \right]} \Psi_\mathrm{O}
    .
\end{eqnarray}

In quantum chemical calculations of electronic structure, molecular orbitals are typically represented as a linear combination of atomic orbitals (LCAO). In practice, the latter are often approximated by a truncated basis set of Gaussian primitives. This approach enables the direct incorporation of the wave function behavior described by Eq.~(\ref{psi}) into the atomic orbitals, resulting in a field- and gauge-dependent orbital centered at the point defined by the vector $\mathbf{K}$
\begin{eqnarray}
    \label{ao}
    \omega_{lm}(\mathbf{r}_{\mathrm{K}}, \mathbf{B}, \mathbf{G}) = \exp \left[ \frac{ \mathrm{i}}{2} \mathbf{B} \times (\mathbf{G} - \mathbf{K}) \cdot \mathbf{r} \right]
    \chi_{lm}(\mathbf{r}_{\mathrm{K}}).\,\,\,
\end{eqnarray}
Here $ \chi_{lm}(\mathbf{r}_{\mathrm{K}})$ refers to the ordinary atomic orbital with quantum numbers $lm$ centered at the origin of the coordinate system, which we may denote as $ \mathbf{K}$. Equation~(\ref{ao}) is often referred to as the London orbitals, a designation that was first proposed by Fritz London \cite{London1937}. Alternatively, it is also known as the GIAOs, which stands for gauge-origin including atomic orbitals or gauge-origin independent atomic orbitals.

In summary, the Hamiltonian presented by Eq.~(\ref{1}) differs from the ordinary field-free Hamiltonian, $ \hat{H}_0 $. It exhibits gauge origin dependence and incorporates the electronic angular momentum operator, leading to the generation of complex wave functions.  Furthermore, the utilization of orbitals in the form of Eq.~(\ref{ao}) also results in the modification of the integrals employed in the most commonly used quantum chemistry libraries (see \cite{Irons2017}, for the derivation of explicit equations). 
Additionally, in the presence of strong magnetic fields, the Born-Oppenheimer approximation requires modification to include the diagonal terms of the non-adiabatic couplings, which account for electron screening of nuclear charges and introduce significant dynamical corrections, particularly at large internuclear distances and in the dissociation limit. Off-diagonal non-adiabatic couplings also become more relevant due to symmetry lowering, especially in cases of near-degenerate electronic states or large-amplitude motion. Nevertheless, the ordinary Born-Oppenheimer approximation remains widely used because it provides a computationally efficient and accurate framework for systems where non-adiabatic effects are minimal, and serves as a practical starting point.

In considering the issue of electronic structure in the presence of a magnetic field, it is evident that the derivation of the coupled-cluster (CC) energy and amplitude equations follows the same procedure as in the absence of a magnetic field, resulting in analogous final equations \cite{Stanton1991}. While the final CC expressions resemble those in the field-free scenario, there is less permutational symmetry for the two-electron integrals \cite{Tellgren_2014}. Furthermore, in the presence of a magnetic field, all related quantities, including integrals, acquire complex values. Consequently, the computational cost is increased, as for each quantity, the real and imaginary parts should be stored, and the operation count for multiplications is four times higher. 

\section{Implementation}
\label{h2}

Our implementation is built upon the Python interface of the GQCP library, which facilitates Hartree-Fock (HF) calculations utilizing gauge-including atomic orbitals (GIAO) in the presence of an external magnetic field \cite{Lemmens2022}. To accommodate both closed- and open-shell molecular species, we employ both restricted Hartree-Fock (RHF) and unrestricted Hartree-Fock (UHF) formalisms. These approaches are used to construct key components for the HF theory, including the Fock matrix \textbf{F}, the overlap matrix \textbf{S}, and a set of electron repulsion integrals, as detailed in the GQCP documentation \cite{GQCP, Lemmens2022}. 

Following this construction, the generalized eigenvalue problem associated with the Hartree-Fock-Roothaan equations is subsequently addressed \cite{Crawford2000}. This step is pivotal for determining the molecular orbital coefficients that minimize the HF energy within the specified basis set. Subsequently, the generalized eigenvalue problemfor Hartree-Fock-Roothaan equations
\begin{eqnarray}
\label{HFR}
    \mathbf{F}\mathbf{C} = \bm{\varepsilon}\textbf{S}\textbf{C},
\end{eqnarray}
is solved in a self-consistent way until the convergence reaches \cite{Crawford2000}. In Eq.~(\ref{HFR}) $\bm{\varepsilon}$ is the diagonal matrix with energies of molecular orbitals and \textbf{C} is the matrix of molecular coefficients. All matrices except the energies of molecular orbitals, in contrast to the zero-field case, can be complex since GIAOs are used and the Hamiltonian, Eq.~(\ref{1}), includes the angular momentum operator. 

Upon obtaining the total energy of the system at the Hartree-Fock level using the GQCP library and generating an array of electron repulsion integrals (ERIs), the CCSD(T) procedure, Coupled Cluster with Single and Double excitations, augmented with perturbative Triple excitations, can subsequently be carried out. This advanced correlation method builds upon the HF reference to provide a more accurate description of electron correlation effects, thereby enhancing the overall accuracy of the computed molecular properties. The ERIs serve as essential inputs for constructing the amplitude equations that define the CCSD(T) formalism, enabling the treatment of both dynamic and non-dynamic correlation effects in a systematic manner.

We used both a self-written algorithm following the basic formulas of the work \cite{Stanton1991}, in which we implemented complex arithmetic, and modified functions of \texttt{PySCF} library. 
It should be noted that the PySCF built-in routines for the SCF procedures are already adapted for complex arithmetic and contain all necessary conjugations for correct evaluations. In the main code, one only needs to comment out the check blocks for intermediate quantities to manage complexity. After this step, the Hartree-Fock results at finite magnetic fields are completely identical to those obtained with \texttt{ChronusQ}, \texttt{BAGEL}, or built-in complex SCF method in \texttt{GQCP}. The same applies to the coupled cluster procedures used.

Using the \texttt{PySCF} built-in tools also allows one to construct one-particle density matrices and perform EOM-CCSD (Equation of Motion Coupled-Cluster with Single and Double excitations) calculation of excited states. The code has been validated in the following way: at zero fields, SCF and CCSD(T) results are compared with calculations performed with \texttt{NWChem} (North-West Computational Chemistry) package \cite{nwchem}. At finite magnetic fields, the SCF energies and permanent dipole moments calculated with the corresponding density matrices are compared with those obtained with \texttt{ChronusQ}. For the cases where CCSD is equivalent to FCI, the results for correlation corrections were compared with those obtained with \texttt{BAGEL} code \cite{BAGEL}. All electronic structure calculations were performed using Dunning's correlation-consistent cc-PVTZ basis sets, modified to account for the gauge-dependent phase as described in Eq.~(\ref{ao}).

\section{Application to light diatomic molecules}
\label{hh}
\subsection{H$_2$ molecule}
\label{h3}

As a first simplest system, we study molecular hydrogen exposed to a strong magnetic field. Earlier such calculations were carried out both within the CCSD and variational approaches with excellent consistency of results between each other \cite{Stopkowicz_2015, Kravchenko}. 

\begin{figure}[hbtp]
    \centering
    \includegraphics[width=0.9\columnwidth]{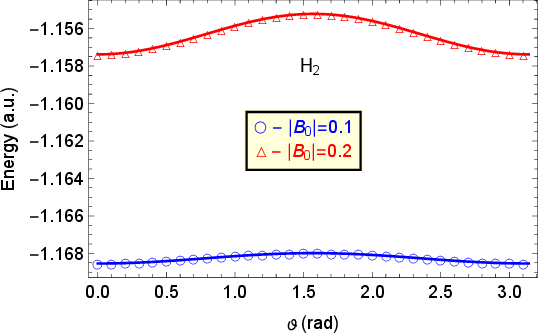}
    \caption{The total CCSD energy of the X$^1\Sigma_{g}^{+}$ state of the H$_2$ molecule as a function of the angle $\theta$ between the magnetic field vector and the molecular axis at different field strengths $B_{0}$ (in atomic units) and fixed geometry. For each value of $B_{0}$, a corresponding equilibrium distance is used, see Table~\ref{tab:req}.}
    \label{fig:1}
\end{figure}

For diatomic molecules, the parallel orientation in the field is generally more favorable. The dependence of the total energy calculated for the equilibrium bond length at a given field strength within the CCSD framework on the angle $\theta$, which determines the direction of the magnetic field relative to the molecular axis, is shown in Fig.~\ref{fig:1}.

From Fig.~\ref{fig:1} one can see that for X$^1\Sigma_{g}^{+}$ state the energy reaches its minima for parallel orientation, i.e. at angles $\theta =0$ and $\theta=\pi$. It should also be emphasized that more complex triatomic molecules, like H$_3$ clusters, in high-strength states also prefer to organize linear structures along the direction of magnetic field vector; see, for example, \cite{turb_h3}. Therefore, in most cases considered below, we will limit ourselves to studying only the orientation of the magnetic field parallel to the axis of the molecule, $B^{\parallel}_{0}$, except for special cases where the perpendicular orientation leads to the formation of bound states. Even though the magnetic field breaks the spatial symmetry of the system, we will continue to use the standard (zero-field) notation for molecular terms for the sake of clarity.

%%%%%%%%

\begin{figure}[hbtp]
    \centering
    \includegraphics[width=0.8\columnwidth]{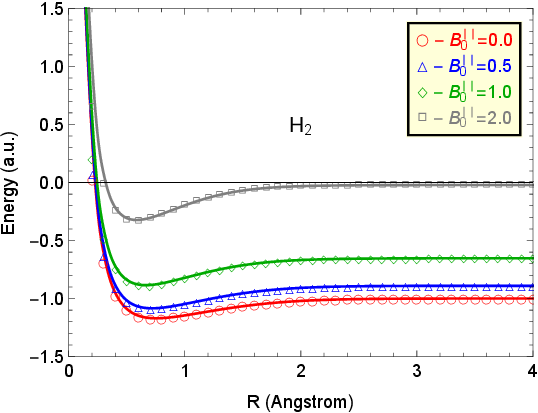}
    \caption{Potential energy curves for the X$^1\Sigma_{g}^{+}$ state of H$_2$
 molecule at different field strength $B^{\parallel}_{0}$ (in atomic units),  oriented parallel to the molecular axis.}
    \label{fig:2}
\end{figure}

As a next step forward in our studies, Fig.~\ref{fig:2} shows the calculated potential energy curves (PECs) for molecular hydrogen H$_2$ as a function of the applied external magnetic field.
The corresponding electron energy is given in atomic units (a.u.) and the internuclear distance in Angstrom, $\AA$. First, the potential energy curve for zero magnetic field is examined. Then the curves for magnetic field strengths $B^{\parallel}_{0} = 0.5 $, $1.0 $ and $2.0 \, \text{a.u.} $ are plotted. These fields are classified as strong, when the magnetic interaction becomes comparable in magnitude to the Coulomb interaction. As shown in Fig.~\ref{fig:2}, the ground state potential energy of a diatomic hydrogen molecule deepens and straightens with increasing field strength, becoming flatter as a function of the internuclear distance.

\begin{figure}[hbtp]
    \centering
    \includegraphics[width=0.4\textwidth]{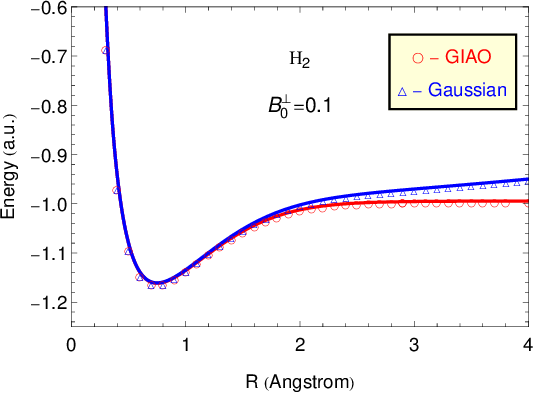}
    \caption{Potential energy curves for the X$^1\Sigma_{g}^{+}$ state of the H$_2$ molecule, calculated with GIAOs and net Gaussian orbitals at magnetic field strength $B^{\perp}_0=0.1$ a.u. perpendicular to the molecular axis}.
    \label{fig:3}
\end{figure}

\begin{table}[h!]
    \centering
    \caption{Equilibrium bond distances $r_{\mathrm{eq}}$ (in Angstrom) for the X$^1\Sigma^{+}_{g}$ state of H$_2$, the X$^1\Sigma^{+}$ states of HeH$^+$ and LiH as a function of magnetic field strength $B^{\parallel}_{0}$.}
    \begin{tabular}{ c c  c  c  c  c}
      \hline
%      \backslashbox{Mol. \kern-2em}{\kern-1em $B^{\parallel}_{0}$, a.u.} & 0.0  & 0.2 & 0.5 & 1.0 & 2.0\\
    $B^{\parallel}_{0}$, a.u. & 0.0 & 0.1 & 0.2 & 0.5 & 1.0 \\
      \hline  
       
         H$_2$   & 0.743 & 0.741 & 0.735 &  0.705 & 0.651 \\
      
         HeH$^+$ & 0.777 & 0.776 & 0.774 & 0.760  & 0.721 \\
      
         LiH     & 1.596 & 1.586 & 1.565 & 1.484 & 1.363 \\
      \hline
    \end{tabular}
    \label{tab:req}
\end{table}
The relevance of using GIAOs in the case of a strong magnetic field is emphasized in Fig.~\ref{fig:3}.

In particular, Fig.~\ref{fig:3} clearly demonstrates that the potential energy curve for the X$^1\Sigma_{g}^{+}$ state of the H$_2$ molecule exhibits different asymptotic behavior at large interatomic distances depending on the basis set used. When employing the net Gaussian basis set (represented by triangles and blue in the online version), the dissociation limit of the H$_2$ molecule is not achieved at perpendicular field orientation, as illustrated for the field strength $B^{\perp}_{0}=0.1$ a.u. In turn, modifying the basis set by including the gauge-origin factor corrects the result, bringing it closer to the asymptotic value of the molecule's dissociation energy. This essentially different behavior of PEC at parallel and perpendicular field directions is rather evident when considering the phase factor in GIAOs: it is zero at parallel orientation and maximal at perpendicular orientation.

\begin{figure*}[htbp]
    \centering
    \includegraphics[width=0.44\textwidth]{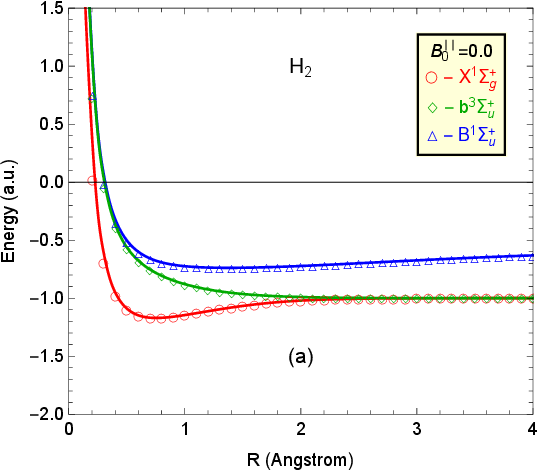}
    \label{fig:4a}
    \includegraphics[width=0.44\textwidth]{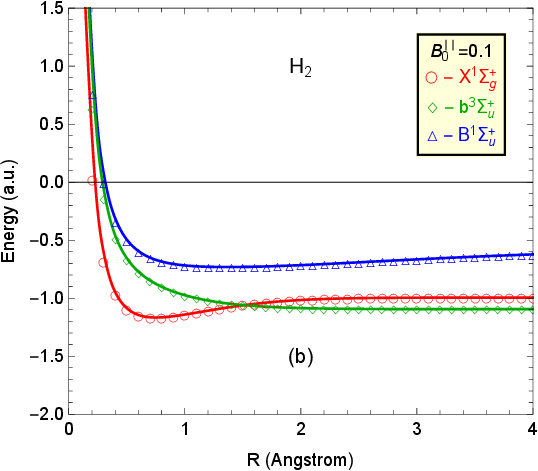}
    \label{fig:4b}
    \includegraphics[width=0.44\textwidth]{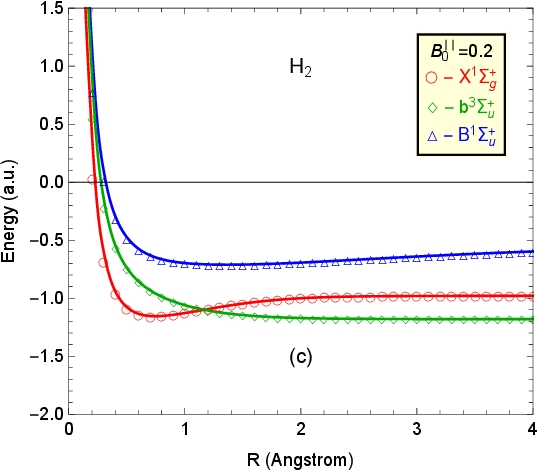}
    \label{fig:4c}
    \includegraphics[width=0.44\textwidth]{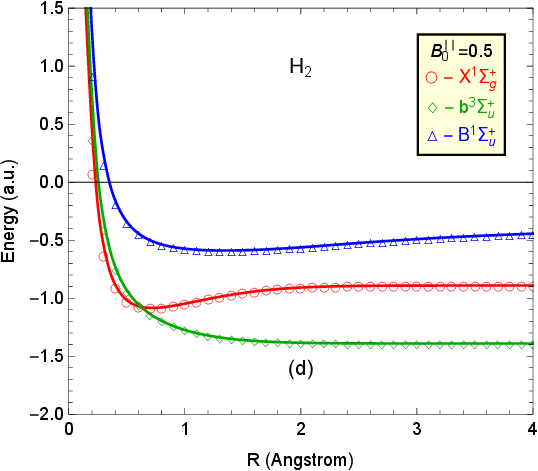}
    \label{fig:4d}
    \includegraphics[width=0.44\textwidth]{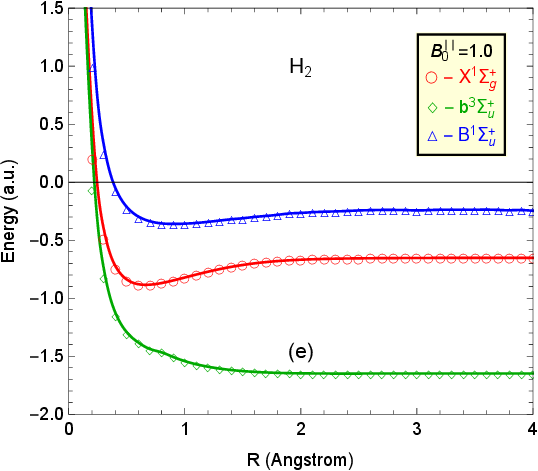}
    \label{fig:4e}
    \caption{Potential energy curves for the X$^1\Sigma_{g}^{+}$, b$^3\Sigma_{u}^{+}$, and B$^1\Sigma_{u}^{+}$ states of the H$_2$ molecule at different field strengths $B_{0}^{\parallel}$ (in atomic units)}
    \label{fig:4}
\end{figure*}

The potential energy curves (PECs) obtained for the ground, first excited singlet, and lowest triplet states are presented in Fig.~\ref{fig:4}. As Fig.~\ref{fig:4} shows, at field strengths above $0.1$ a.u., the X$^1\Sigma_{g}^{+}$ state is no longer the lowest energy state and is replaced by a non-bonding triplet state with a repulsive character. This confirms earlier calculations which indicated that bound states are no longer formed in the magnetic fields considered~\cite{Kravchenko, Kravchenko2, Nader2021}. 
Additionally, as shown clearly in Fig.~\ref{fig:4}, the introduction of a magnetic field lifts the degeneracy between the singlet and triplet states in the dissociation limit, resulting in distinct energy levels for the two separated hydrogen atoms.

\begin{figure}[hbtp]
    \centering
    \includegraphics[width=0.48\textwidth]{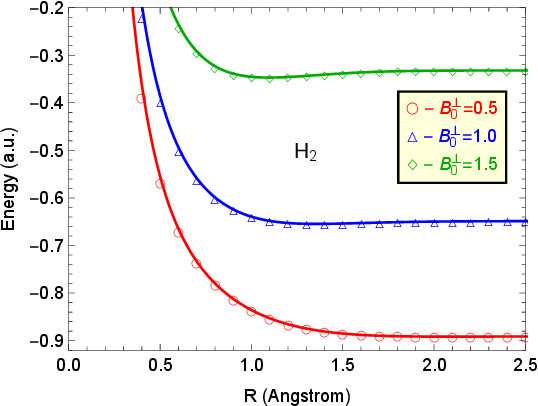}
    \caption{Comparison of potential energy curves for the b$^3\Sigma_{u}^{+}$ state of the H$_2$ molecule at different magnetic field strength, oriented perpendicular to the molecular axis. Starting from the values $B^{\perp}_{0}$ greater than 1 a.u., the curves exhibit the emergence of a potential well, which distinctly arises from the mechanism of perpendicular paramagnetic bonding discussed in \cite{Lange2012}.}
    \label{fig:5}
\end{figure}

It should be noted that, taking into account the GIAO orbitals, the dissociation limit of the H$_2$ molecule is neatly recovered for all the fields mentioned. To emphasise the importance of the approach described in section~\ref{h1}, the calculation results for the field perpendicular to the molecular axis are shown in Fig.~\ref{fig:5}.

In a magnetic field starting from a strength of $0.5$ a.u., the lowest triplet term does not form a bound molecular state and remains purely repulsive. In particular, the corresponding potential curve lacks a well-defined minimum and therefore there is no equilibrium internuclear distance in the triplet molecular term for a parallel field orientation. The $1.0$ a.u. field represents a transitional case where the triplet state can transfer to the singlet term of a hydrogen molecule as the nuclei approach. In this scenario the total spin momentum is rearranged in a radiationless process to form a singlet state. For the perpendicular orientation of the field with a strength greater than $1.0$ a.u., see Fig.~\ref{fig:5}, the triplet state is bound via a chemical bonding mechanism that only occurs in external magnetic fields, called paramagnetic bonding \cite{Lange2012}. For magnetic fields greater than $12.3$ a.u. the molecule reappears as a stable system in $^3\Pi_{u}$ with total spin projection $M_{s}=-1$ towards the magnetic field direction, see \cite{Kravchenko}. 

\begin{figure}[hbtp]
    \centering
    \includegraphics[width=0.55\textwidth]{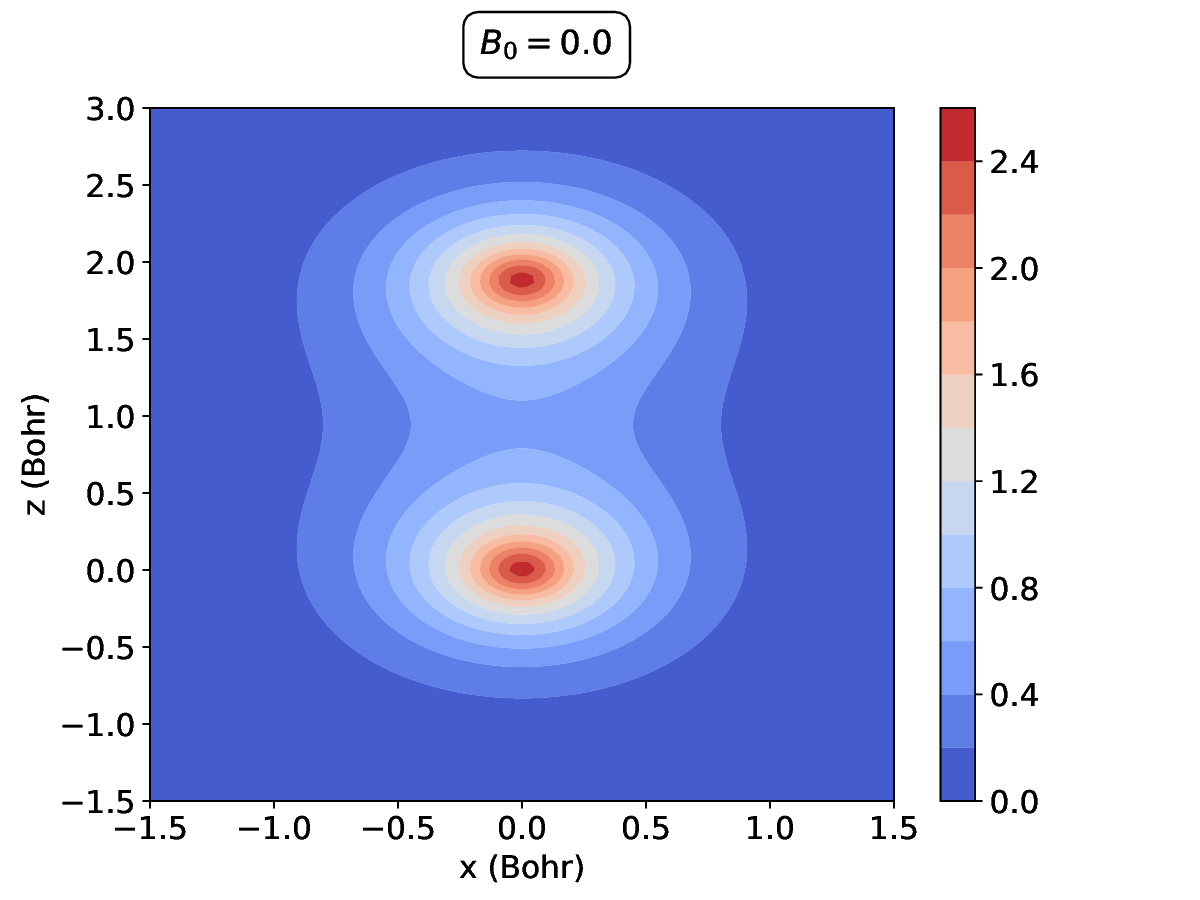}
    \includegraphics[width=0.55\textwidth]{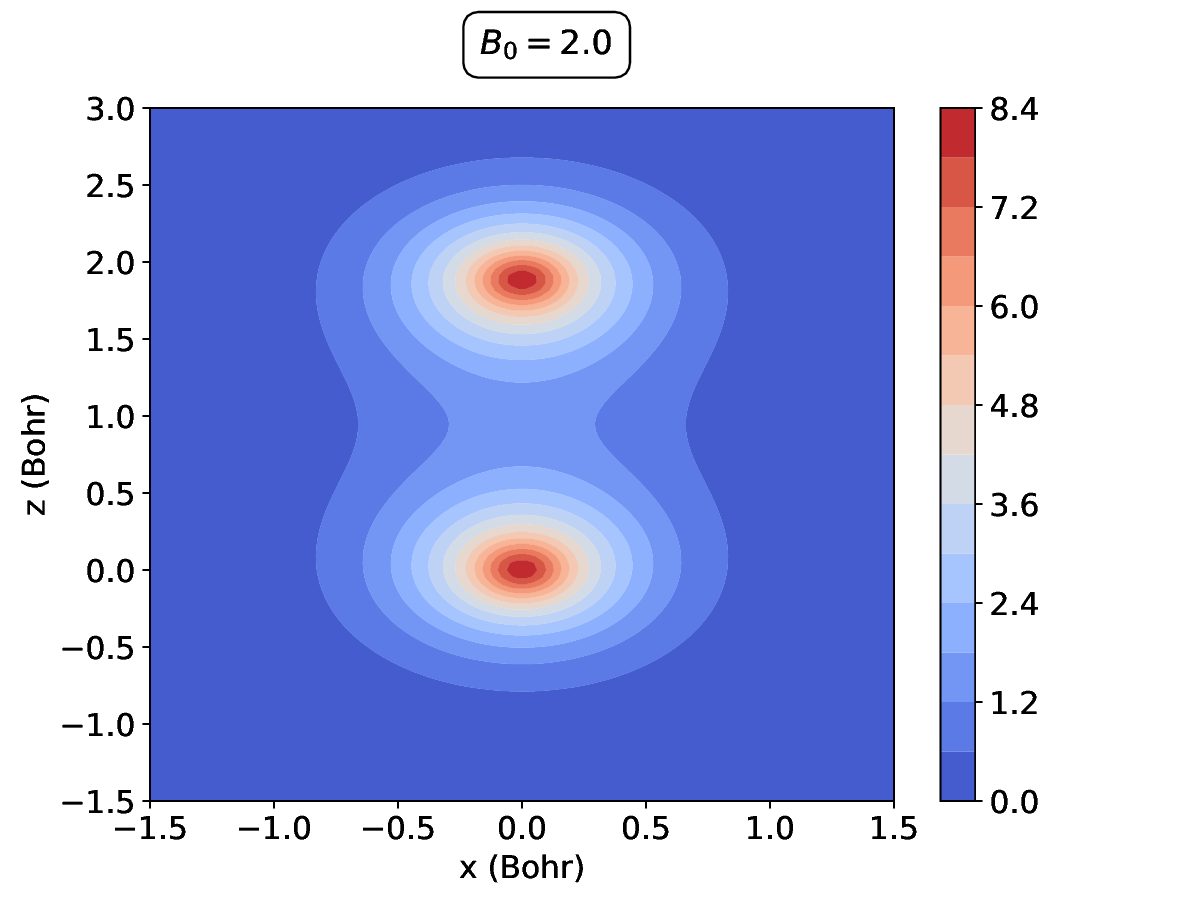}
    \includegraphics[width=0.55\textwidth]{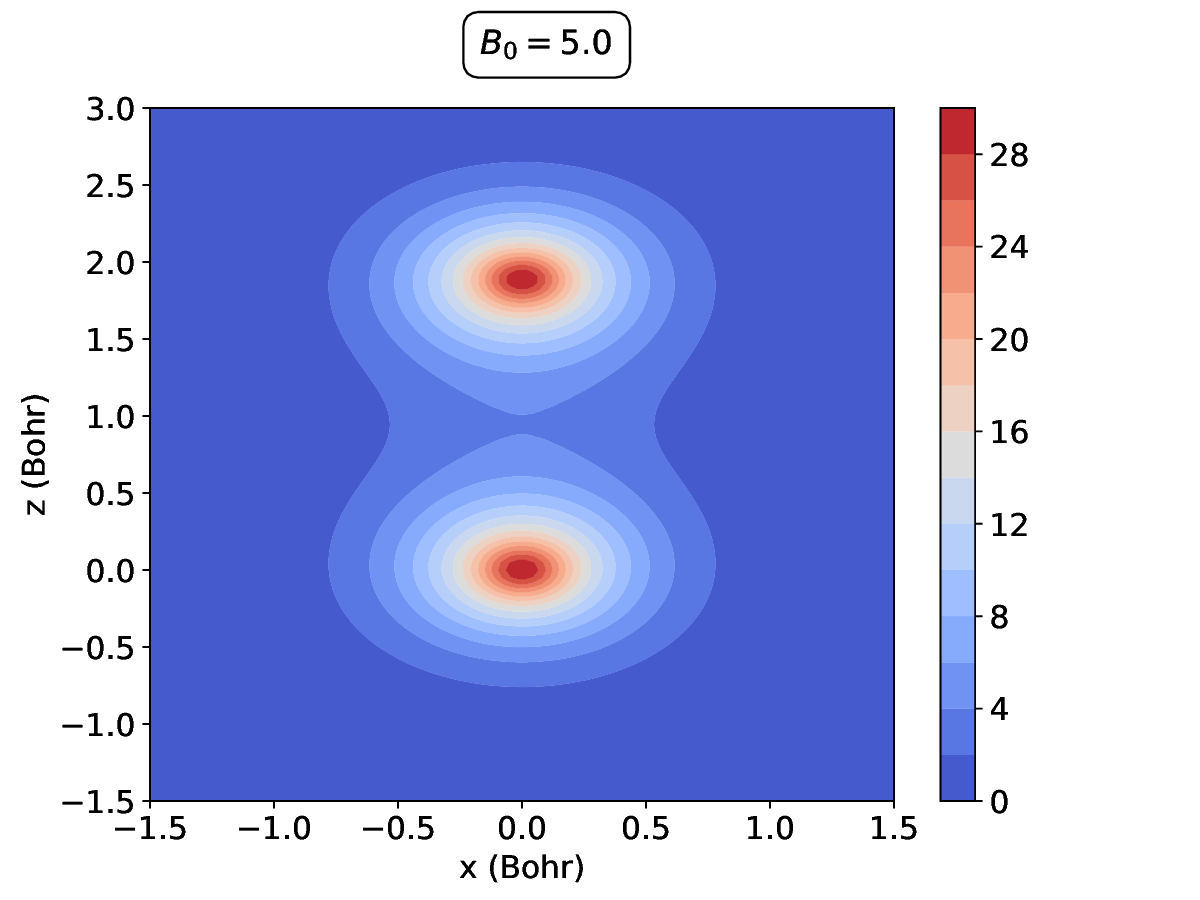}
    \caption{The contour plot of the electron charge density of H$_2$ molecule in its ground state X$^1\Sigma_{g}^{+}$ at different field strength $B^{\parallel}_{0}$ (in atomic units). For all plots, the distance between the nuclei was kept constant at $1$ Angstrom.}
    \label{fig:7}
\end{figure}

It may also be instructive to examine the behavior of the charge density in the presence of an external magnetic field. For this purpose, the charge density at zero field and $B^{\parallel}_{0}=2.0$, $B^{\parallel}_{0}=5.0$ a.u. field strengths (oriented along the molecular axis) is shown in Figs.~\ref{fig:7}, providing a visual representation of the charge density at a fixed interatomic distance $1$ \AA. The inclusion of a magnetic field significantly increases the localisation of the electron density along the direction parallel to the applied field, while simultaneously extending it in the direction perpendicular to the field. This is also accompanied by a known decrease in the equilibrium distance compared to the field-free case
\cite{Kravchenko}. 

Since the hydrogen molecule is homonuclear, it has no dipole moment in X$^1\Sigma_{g}^{+}$ state. The study of this property for heteronuclear species like HeH$^+$ and LiH is discussed in the following sections. Another important characteristic is the transition dipole moments (TDMs) between singlet states, as they govern the cross-sections of inelastic photo-association and radiative charge exchange \cite{Stancil1993, Loreau2013}. The TDMs are directly incorporated into the corresponding rate equations, thereby enabling the estimation of molecular concentrations and line intensities. In the absence of magnetic fields, such calculations have been routinely performed since the 1990s \cite{Zygelman1989, Zygelman1998, Forrey2020}. The accurate computation of TDMs is crucial for modeling radiative processes in astrophysical environments, such as stellar atmospheres and interstellar clouds, where these processes influence molecular formation and spectral observations.

\begin{figure}[hbtp]
    \centering
    \includegraphics[width=0.44\textwidth]{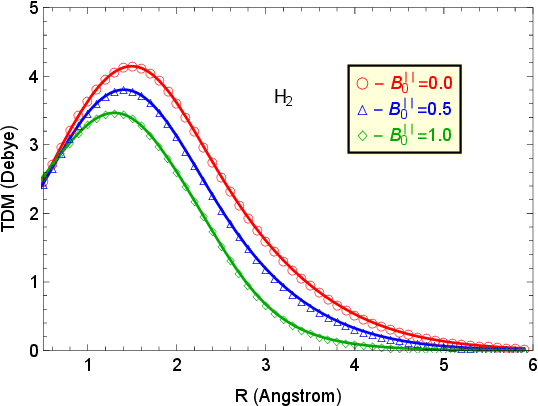}
    \caption{Transition dipole moment (TDM) between the X$^1\Sigma^{+}_{g}$ and B$^1\Sigma^{+}_{u}$ states of the H$_2$ molecule, plotted as a function of internuclear distance for various field strengths $B^{\parallel}_{0}$ (in atomic units). }
    \label{fig:6}
\end{figure}

The transition dipole moment (TDM) between the ground state and the first excited singlet in an external magnetic field, taking into account correlation effects, can be calculated using PySCF built-in routines. The PySCF library provides tools for the calculation of off-diagonal dipole moments between arbitrary states (or TDM determination). This is achieved by applying the Configuration Interaction with Single and Doubles (CISD) method, which relies on a specific Hartree-Fock reference. 
The corresponding results for TDMs for different field strengths as a function of the internuclear distance are shown in Fig.~\ref{fig:6} for the parallel field configuration.

\subsection{HeH$^{+}$ molecule}
\label{HeH+}

The HeH$^{+}$ molecule holds great significance in astrophysics, as it is believed to be the first molecular bond formed in the universe. In the aftermath of the Big Bang, as the universe began to cool, helium and hydrogen combined to form HeH$^{+}$, marking a pivotal moment in the chemical evolution of the cosmos. This molecule played a crucial role in shaping the chemistry of the early universe and influencing the formation of subsequent chemical elements and molecules. The detection of HeH$^{+}$ in space provides valuable insights into the conditions of the early universe, allowing the refinement of models of cosmic evolution and star formation. Furthermore, recent observations in the NGC 7027 planetary nebula \cite{G_sten_2019} have underscored the astrophysical importance of this molecule, sparking renewed interest in its study.

\begin{figure}[hbtp]
    \centering
    \includegraphics[width=\columnwidth]{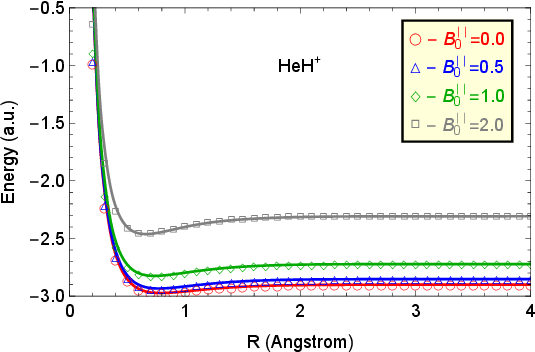}
    \caption{Potential energy curves for the X$^1\Sigma^{+}$ state of HeH$^+$
 molecule at different field strength $B{^{\parallel}}_{0}$ (in atomic units). }
    \label{fig:8}
\end{figure}

Therefore, the behavior of the HeH$^{+}$ molecular ion in superstrong magnetic fields needs to be carefully studied; see the first theoretical studies \cite{Heyl_1998, Turbiner_2007}. In the singly ionized HeH$^{+}$ molecule, the electrons are predominantly localized on the helium atom as a result of the stronger Coulomb interaction between the electrons and the nucleus compared to that in a neutral hydrogen molecule. As a consequence, the influence of an external magnetic field is substantially diminished. The evolution of the ground-state potential curve as a function of increasing magnetic field strength is depicted in Fig.~\ref{fig:8}.
It can be seen that the PECs for the $ X^1\Sigma^{+}$ state of the HeH$^+$ molecule, calculated at various external magnetic field strengths of $ 0.0 $, $0.8$, $ 1.0 $, and $ 2.0 $ atomic units, exhibit a behavior that closely mirrors the PEC characteristics observed in the hydrogen molecule H$_2$. 

\begin{figure*}[htbp]
    \centering
    \includegraphics[width=0.9\columnwidth]{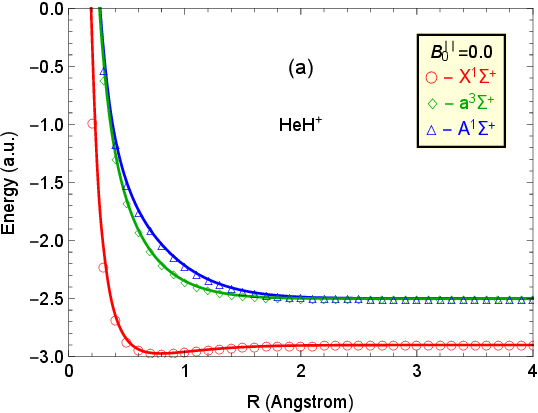}
    \label{fig:9a}
    \includegraphics[width=0.9\columnwidth]{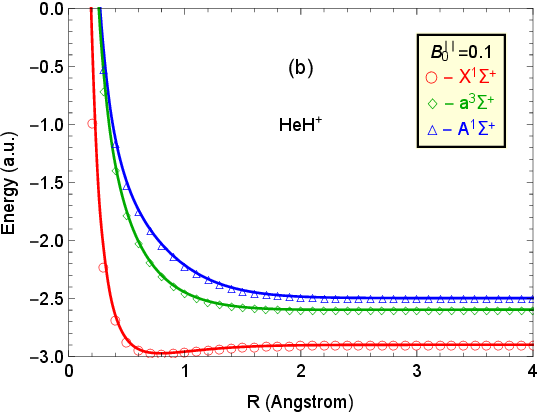}
    \label{fig:9b}
    \includegraphics[width=0.9\columnwidth]{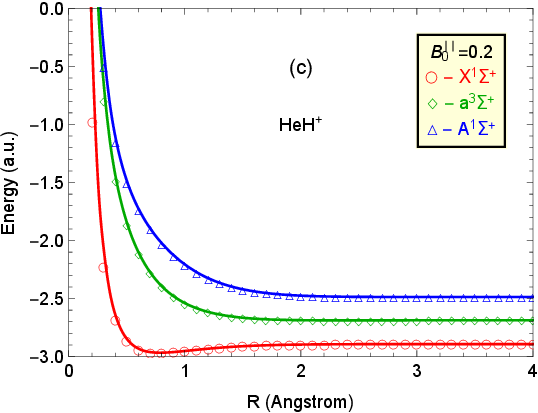}
    \label{fig:9c}
    \includegraphics[width=0.9\columnwidth]{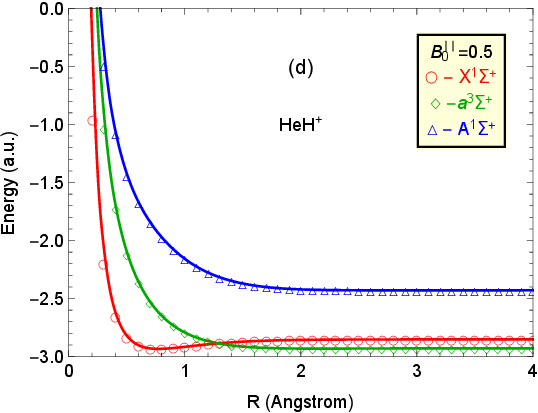}
    \label{fig:9d}
    \includegraphics[width=0.9\columnwidth]{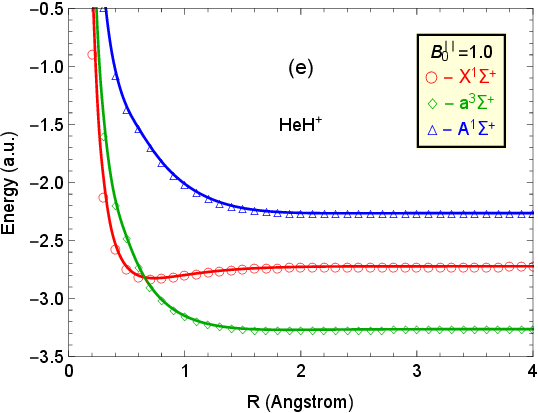}
    \label{fig:9e}
    \caption{Potential energy curves for the X$^1\Sigma^{+}$, a$^3\Sigma^{+}$, and A$^1\Sigma^{+}$ states of the HeH$^{+}$ molecule, at different field strengths $B^{\parallel}_{0}$ (in atomic units).}
    \label{fig:9}
\end{figure*}

Following a similar approach to that used for the hydrogen molecule (Section~\ref{h2}), we have constructed potential energy curves for the ground state X$^1\Sigma^{+}$, the excited singlet state A$^1\Sigma^{+}$, and the triplet state a$^3\Sigma^{+}$ of HeH$^{+}$. The corresponding graphs are presented in Fig.~\ref{fig:9}. Notably, at magnetic field strengths exceeding $0.5$ a.u., the triplet state emerges as the lowest unbound state.

\begin{figure}[hbtp]
    \centering
    \includegraphics[width=0.44\textwidth]{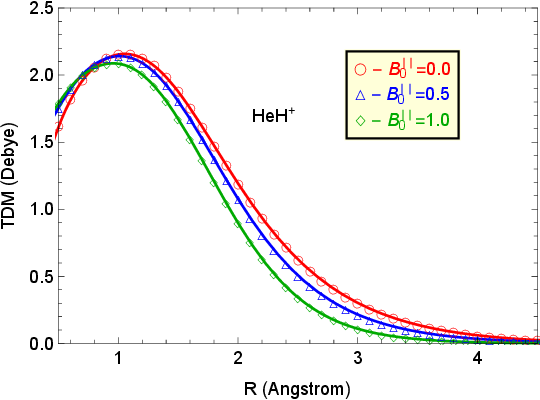}
    \caption{Transition dipole moment (TDM) between the X$^1\Sigma^{+}$ and A$^1\Sigma^{+}$ states of the HeH$^+$ molecule, plotted as a function of internuclear distance for various field strengths $B^{\parallel}_{0}$ (in atomic units). }
    \label{fig:10}
\end{figure}

Using the CISD procedure, we have calculated the TDM between the X$^1\Sigma^{+}$ and A$^1\Sigma^{+}$ states of the HeH$^+$ molecule, with a focus on the dependence on the internuclear distance and the magnetic field strength $B^{\parallel}_{0}$. The results of these calculations are presented in Fig.~\ref{fig:10}.
The zero-field transition dipole moments (TDMs) of HeH$^+$, presented in Fig.~\ref{fig:10}, agree excellently with those reported in \cite{Zygelman1989, Zygelman1998}. Unlike the TDMs of the H$_2$ molecule, which vary significantly with magnetic field strength, those of HeH$^+$ show only minor changes while maintaining the same dependence on internuclear distance. This reduced sensitivity to the magnetic field arises from the stronger Coulomb interaction due to the greater localization of electrons around the helium nucleus.

In contrast to the transition dipole moments, the total permanent dipole moment (PDM) $\bm{\mu}_{\text{total}} $ of the HeH$^+$ molecular ion in X$^1\Sigma^{+}$ state remains almost constant with respect to the field strength. This is evident from the corresponding definition of PDM:
\begin{eqnarray}
\label{tot_dip}
\bm{\mu}_{\text{total}} = -e \int \mathbf{r} \, \rho(\mathbf{r}) \, d\mathbf{r} + e\sum_k Z_k  \, \mathbf{R}_k,
\end{eqnarray}
where $\rho(\mathbf{r})$ is the electron density, $Z_k$ is the charge of the $k$-th nucleus, and $\mathbf{R}_k$ is the vector from the coordinate origin (center of mass) to the corresponding nucleus, the electron charge $e$ is written out for clarity.

\begin{figure}[hbtp]
    \centering
    \includegraphics[width=\columnwidth]{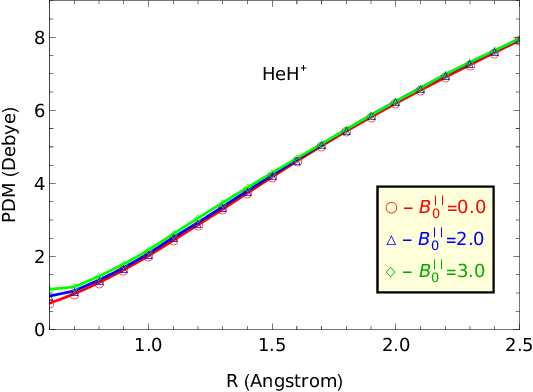}
    \caption{Permanent dipole moment (PDM) for HeH$^+$ molecule (in Debye) in X$^1\Sigma^{+}$ state as a function of internuclear distance for various field strengths $B^{\parallel}_{0}$ (in atomic units).}
    \label{fig:11}
\end{figure}

In the case of the X$^1\Sigma^{+}$ state, all the electrons are localized at the nucleus of the helium atom. Therefore the contribution to the permanent dipole moment is determined by the second term and increases linearly with increasing internuclear distance (this term depends on the choice of coordinate system for charged molecules). The corresponding curve of the PDM for HeH$^+$ is shown in Fig.~\ref{fig:11}. 

It is evident that at large distances, the PDM shows negligible dependence on the field strength. In this region, the total value is primarily determined by the nuclear component, as described by the second term in Eq.~(\ref{tot_dip}). In contrast, at short distances, the electronic component has a more significant effect, which is reflected in the bending of the straight line. This influence also depends on the strength of the field. However, it can be concluded that the effect of the electronic component is largely insignificant within the range of internuclear distances considered.

\subsection{LiH molecule}
\label{LiH}

Lithium hydride (LiH) has been the subject of extensive theoretical and experimental studies in molecular physics over the years \cite{Partridge1981}. As one of the simplest neutral heteronuclear diatomic molecules, LiH represents a critical benchmark system to assess the precision of diverse quantum chemical methodologies \cite{Tung2011, Holka2011}. In the context of astrophysics, LiH is significant because of its potential role in the cooling of primordial gas clouds \cite{Bovino2011}, which is essential to understand the formation of the first stars and galaxies. Furthermore, LiH provides valuable information on the evolution of stars and interstellar clouds \cite{Dulick1998}. The detection of lithium hydride and other lithium-containing molecules, such as LiCl and LiOH, in the atmospheres of dwarf stars is also of great interest \cite{Weck2004}. Despite the potential applications of LiH in astrophysics, searches for this molecule at high redshift have not yet yielded positive results \cite{Friedel2011}. It is worth noting that for calculations of the concentration of this molecule in the interstellar medium, as well as in the vicinity of astrophysical sources, it is also necessary to know the dipole moment as a function of the internuclear distance; see \cite{Dalgarno1996}.

\begin{figure}[hbtp]
    \centering
    \includegraphics[width=\columnwidth]{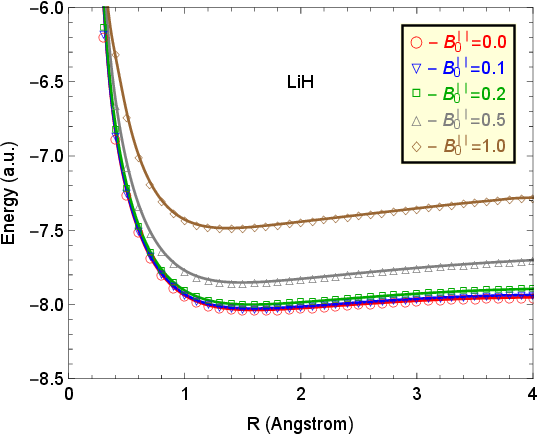}
    \caption{Potential energy curves for the X$^1\Sigma^{+}$ state of the LiH molecule at different magnetic field strengths $B^{\parallel}_{0}$ (in atomic units).}
    \label{fig:12}
\end{figure}

\begin{figure*}[hbtp]
    \centering
    \includegraphics[width=\columnwidth]{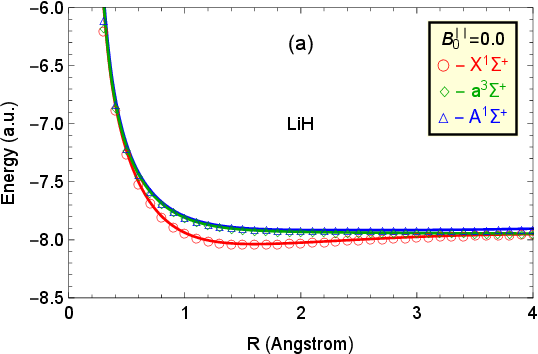}
    \label{fig:9a}
    \includegraphics[width=\columnwidth]{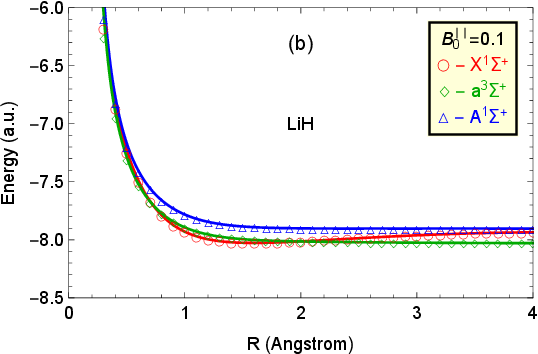}
    \label{fig:9b}
    \includegraphics[width=\columnwidth]{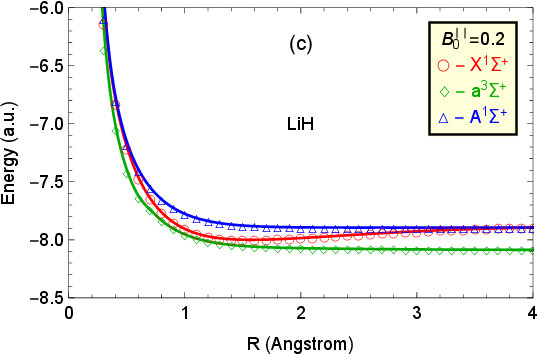}
    \label{fig:9c}
    \includegraphics[width=\columnwidth]{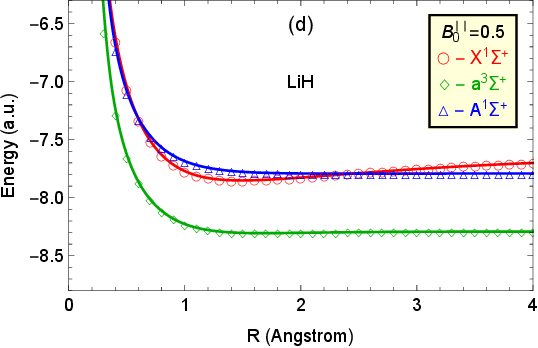}
    \label{fig:9d}
    \includegraphics[width=\columnwidth]{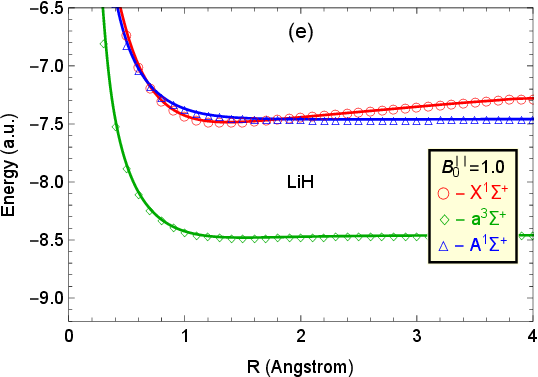}
    \label{fig:9e}
    \caption{Potential energy curves for the X$^1\Sigma^{+}$, a$^3\Sigma^{+}$, and A$^1\Sigma^{+}$ states of the LiH molecule, at different field strengths $B^{\parallel}_{0}$ (in atomic units).}
    \label{fig:13}
\end{figure*}

In the following, we re-examine the potential energy curves for the ground and excited states of LiH at various magnetic field strengths. The results of numerical calculations are presented in Figs.~\ref{fig:12},~\ref{fig:13}, see, e.g., Ref.~\cite{Monzel2022} for comparison. For a more detailed analysis as before, the constructed potential curves of the X$^1\Sigma^{+}$ term for different magnetic field strengths are shown in a separate Fig.~\ref{fig:12}.
The ground, X$^1\Sigma^{+}$ and excited a$^3\Sigma^{+}$ and A$^1\Sigma^{+}$ terms are shown in Fig.~\ref{fig:13} for the fields of strengths $0.0$, $0.1$, $0.2$, $0.5$ and $1.0$ a.u., oriented along the molecular axis.

\begin{figure}[hbtp]
    \centering
    \includegraphics[width=\columnwidth]{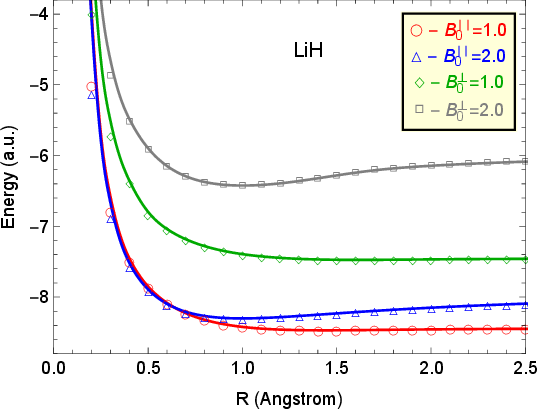}
    \caption{Potential energy curves for the a$^3\Sigma^{+}$ state of the LiH molecule at parallel $B^{\parallel}_{0}$ and perpendicular $B^{\perp}_{0}$ field orientations (in atomic units). The triplet state becomes bound in both cases at field strengths above $1$ a.u. The corresponding equilibrium distances are (in Angstrom): $r_{\mathrm{eq}}(B^{\parallel}_{0}=1.0)=1.502$, $r_{\mathrm{eq}}(B^{\parallel}_{0}=2.0)=0.984$, $r_{\mathrm{eq}}(B^{\perp}_{0}=1.0)=1.607$, $r_{\mathrm{eq}}(B^{\perp}_{0}=2.0)=0.997$.}
    \label{fig:14}
\end{figure}

With regard to molecules H$_2$ and HeH$^+$, the behavior of the potential energy curves as a function of the internuclear distance shows analogous tendencies for LiH. From Fig.~\ref{fig:13} it can be seen that the crossing of the singlet ground molecular term and the triplet state occurs at much lower field values compared to the H$_2$ compound. The repulsive character of the triplet term is preserved up to $B^{\parallel}_{0}=1.0$ a.u. and then becomes strictly binding with increasing field strength even for the parallel configuration, see Fig.\ref{fig:14} for details. 

Thus, in contrast to the hydrogen molecule, the first triplet state of LiH turns out to be binding not only for the perpendicular field orientation. Moreover, as can be seen from the figure \ref{fig:14}, the parallel orientation is energetically more favourable (because it is lower). It is worth noting that the formation of triplet bound states in LiH is visible even at the Hartree-Fock level, i.e. without the inclusion of electron correlations, and can be found just within the existing codes such as \texttt{ChronusQ}.

\begin{figure}[hbtp]
    \centering
    \includegraphics[width=0.45\textwidth]{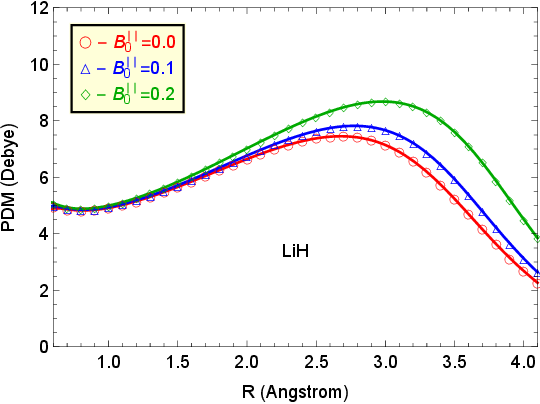}
    \caption{Permanent dipole moment for LiH molecule (in Debye) in X$^1\Sigma^{+}$ state as a function of internuclear distance for various field strengths $B^{\parallel}_{0}$ (in atomic units).}
    \label{fig:15}
\end{figure}

From Fig.~\ref{fig:13} it can be seen that the excited singlet state of the LiH compound in a magnetic field is repulsive (no well is visible on the potential curve). The situation is analogous to the weakly bound molecule HeH$^+$, for which the equilibrium distance of the nuclei exists only for the ground term. In addition, the difference to the HeH$^+$ molecule is that the permanent dipole moment of LiH in its X$^1\Sigma^{+}$ state shows a pronounced sensitivity to the field strength. The results of the PDM calculations are shown in Fig.~\ref{fig:15} (compare with Fig.~\ref{fig:11}). Specifically, the permanent dipole moment curve is non-linear with internuclear distance, albeit retaining its structure up to field strength $B^{\parallel}_{0}=0.2$ a.u.

\subsection{Calculations of equilibrium distances and comparison total energies and PDMs of H$_2$, HeH$^+$ and LiH molecules}
\label{req}

In this section, we present the results of numerical calculations of equilibrium distances, $r_{\mathrm{eq}}$, and the corresponding potential surface minima. The total energies are compared in the context of two approaches: the \texttt{GQCP+PySCF} and \texttt{BAGEL} packages \cite{GQCP,PySCF,BAGEL}. Calculations are performed both in the absence of a magnetic field and at parallel field strengths $B_0$.

Starting from the equilibrium distances for the diatomic molecules H$_2$, HeH$^+$ and LiH, the results are collected in Table~\ref{tab:req}, where each row shows the values of $r_{\mathrm{eq}}$ for the given molecular compound as a function of the magnetic field strength.

From Table~\ref{tab:req} it is seen that an external magnetic field adds a diamagnetic "squeeze" (from the $\mathbf{A}^2$ term) plus orbital‐magnetic couplings, which compress the electron cloud. This enhanced electronic localization strengthens the bond and shifts the equilibrium bond length to a smaller value. 
The obtained value of the equilibrium distance of molecular hydrogen in the zero field can be compared with the well-established value of $0.741$ \AA, see \cite{NIST} and references therein. The result found within the \texttt{GQCP+PySCF} approach, $0.743$ \AA, agrees with an accuracy better than $0.5$\%, which is satisfactory for further calculation. Comparison of the equilibrium bond distances given for the zero-field helium hydride cation with experimental data \cite{Perry2014} yields excellent agreement for $r_{\mathrm{eq}}$, see also \cite{NIST} for the corresponding calculations under different methods. In turn, the discrepancy $r_{\mathrm{eq}}$ for lithium hydride is close to $0.3\%$ for the zero-field case.

For the strong magnetic field cases of $0.2$, $0.5$ and $1.0$ a.u., the difference between $r_{\mathrm{eq}}$ for molecular hydrogen leads to a relative deviation of our results from the results of \cite{Kravchenko} at the $1\%$ level, see also \cite{Doma2016} for the results found in the variational Monte Carlo method. For the molecular ion HeH$^+$ our results are in agreement with the calculated data presented in \cite{Turbiner_2007} at the level of $1-3\,\%$. Finally, the results for $r_{\mathrm{eq}}$ in the LiH compound in the zero and non-zero field agree to within $1\%$ with the values reported in \cite{Runge1997}.

Based on the calculated equilibrium bond distances, numerical results for the total energies are given in Table~\ref{tab:CCvsCI}. In particular, for a given compound, the first row represents the values obtained within the \texttt{GQCP+PySCF} approach, and the second row corresponds to the relativistic Full-CI calculations performed within the \texttt{BAGEL} code. The calculated energies agree with each other with an accuracy greater than $0.2 \%$. All energies are computed at the corresponding equilibrium distances listed in Table~\ref{tab:req}.
\begin{table}[h]
      \caption{Comparison of the total energies (in atomic units) of the H$_2$ and HeH$^+$ molecules obtained with the \texttt{GQCP+PySCF} approach (first row) at the CCSD level of theory with the values calculated with the \texttt{BAGEL} code (second row) in the framework of the Relativistic Full Configuration Interaction (RelFCI) + GIAO. All energies are calculated at the equilibrium distances given in table~\ref{tab:req}.}
    \begin{tabular}{c c c c c c}
      \hline
       $B^{\parallel}_{0}$, a.u. & 0.0  & 0.1 & 0.2 & 0.5 & 1.0\\
     \hline  
         H$_2$   & -1.17234 & -1.16854 & -1.15738 & -1.08686 & -0.88507\\
                 & -1.17235 & -1.16856 & -1.15740 & -1.08688 & -0.88512\\
                 \hline
         HeH$^+$ & -2.97518 & -2.97357 & -2.96874 & -2.93578 & -2.82713\\
                 & -2.97504 & -2.97343 & -2.96860 & -2.93565 & -2.82702\\ 
         %         \hline
         % LiH     & -8.03662 &  &  &  & \\
         %         &   & &  &  & \\
      \hline
      \end{tabular}   
    \label{tab:CCvsCI}
\end{table}

The agreement of the total energy values with the corresponding data available in the literature is also satisfactory (with respect to the defined deviations for $r_{\mathrm{eq}}$). For example, for molecular hydrogen in the ground state, the deviation from the result of \cite{Kravchenko} in the $1$ a.u. field is $1\%$ and is slightly better for the zero-field case. In the absence of a magnetic field, the total energy of the LiH molecule is in excellent agreement with the results reported in \cite{Hyams1994,Cafiero2002}, with a relative deviation of about $0.2\%$. The relative deviation with respect to recent results \cite{Hampe2019} remains at the same level for strong magnetic field strengths.

Finally, the numerical results for the permanent dipole moment of the LiH molecule obtained by different calculation methods are compared in Table~\ref{tab:pdm_comparison}. At zero field the results are in agreemnt with those obtained in \cite{Dalgarno1996}. Permanent dipole momentum values were obtained for parallel oriented field $B^{\parallel}_{0}=0.0$, $0.1$ and $0.2$ a.u. depending on the internuclear distance. In Table~\ref{tab:pdm_comparison}, the PDM values evaluated with the \texttt{ChronusQ} code (second column) are compared with those obtained with the adapted \texttt{GQCP+PySCF} approach, both at the same SCF level of theory, while the third column, includes correlation corrections to the PDM within the CCSD method. 

\begin{table}[h]
    \centering
    \caption{Comparison of the permanent dipole moment of the X$^1\Sigma^+$ state (in Debye) as a function of internuclear distance $R$ (in Angstrom) for the LiH molecule at various magnetic field strengths, calculated using the \texttt{GQCP+PySCF} and \texttt{ChronusQ} code. The results are presented for different levels of theory, including self-consistent field (SCF) and coupled clusters with singles and doubles (CCSD).}
    \begin{tabular}{c c c c}
      \hline
       & & $B^{\parallel}_{0}=0.0$ a.u. & \\
      \hline
       $R$ & \texttt{ChronusQ} (SCF) & \texttt{GQCP+PySCF} (SCF) & \texttt{GQCP+PySCF} (CCSD)\\
      \hline
       1.0 & 4.965  & 4.965  & 4.910\\
       2.0 & 7.005  & 7.005  & 6.647\\
       3.0 & 9.939  & 9.939  & 7.156\\
       4.0 & 12.743 & 12.743 & 2.675\\
      \hline  
    & & $B^{\parallel}_{0}=0.1$ a.u. & \\
      \hline
       1.0 & 4.983 & 4.983  & 4.938\\
       2.0 & 7.083 & 7.083  & 6.770\\
       3.0 & 10.187& 10.187 & 7.676\\
       4.0 & 13.269 & 13.269 & 3.117\\
     \hline  
    & & $B^{\parallel}_{0}=0.2$ a.u. & \\
      \hline
       1.0 & 5.019 & 5.019 & 4.992  \\
       2.0 & 7.259  & 7.259 & 7.024  \\
       3.0 & 10.668 & 10.668 & 8.690  \\
       4.0 & 14.253 & 14.253 & 4.507 \\   
       \hline
    \end{tabular}
    \label{tab:pdm_comparison}
\end{table}

At the SCF level of theory, \texttt{GQCP+PySCF} shows excellent agreement with results obtained using the \texttt{ChronusQ} package, but differs significantly from the results accounting for correlation effects. The difference grows as the internuclear distance increases both in the absence and in a strong magnetic field. It should be emphasised that the third column in Table~\ref{tab:pdm_comparison}, as the coupled clusters method corrects the PEC to the proper dissociation limit and electron density, ensures the correct behavior. Consequently, the implementation of the coupled clusters method in combination with the gauge-including atomic orbitals developed in this work leads to accurate predictions of the PDM at large distances, in contrast to the single-determinant approach.

\section{Vibrational structure in the presence of magnetic field}
\label{h4}

As the final calculations in this study, we present results for the vibrational structure of the molecular compounds H$_2$, HeH$^+$ and LiH exposed to an external magnetic field. The vibrational-rotational spectroscopic behavior in strong magnetic fields poses a complex and nuanced challenge for quantum chemical modeling. The traditional Born-Oppenheimer (BO) approximation, a cornerstone of quantum chemistry computations, encounters significant theoretical limitations when applied to molecular systems under intense magnetic fields \cite{Avron1978, Avron1978_2, Schmelcher1988}.

The principal difficulty emerges from the non-adiabatic coupling terms induced by the magnetic field, which exhibit a non-linear dependence on the field strength and are absent in zero-field scenarios. Under extreme conditions, these coupling terms become increasingly pronounced, progressively undermining the validity of the standard adiabatic approximation. Specifically, the diagonal non-adiabatic coupling terms, which are typically considered minor corrections in conventional zero-field treatments, can substantially perturb molecular dynamics in strong magnetic regimes \cite{Schmelcher1988}.

In \cite{Schmelcher1988}, the crucial role of incorporating diagonal non-adiabatic coupling terms into the structure of the adiabatic approximation was meticulously delineated. These terms elucidate field-dependent contributions that significantly enhance the precision of nuclear motion descriptions, providing more accurate theoretical calculations. Moreover, it was shown that off-diagonal couplings acquire heightened relevance, especially in molecular collision processes.

However, the calculation of adiabatic corrections in a strong magnetic field is a separate challenge. The accurate calculation of the vibrational structure, taking into account the non-adiabatic  nuclear motion and the corresponding development of the necessary numerical procedures, is of considerable interest, but is left for future work. Being no less important, calculations of the vibrational structure in the BO approximation enable us to trace the evolution of vibrational lines as the magnitude of the field changes. In this approximation, the problem reduces to solving the one-dimensional Schr\"odinger equation for the potential energy curve $V(R)$, calculated at the previous step (see section~\ref{hh}) \cite{Yurchenko_2016}. For parallel orientation  of the magnetic field it reads:
\begin{eqnarray}
\label{vibr}
\left(-\frac{1}{2\mu} \frac{d^2 }{dR^2} + \frac{J(J+1)}{2\mu R^2}+  V(R, \theta = 0)\right) \psi_{\nu J} 
\\\nonumber
= E_{\nu J} \psi_{\nu J},    \qquad
\end{eqnarray}
where $\psi_{\nu J}$ is the wave function corresponding to the state with  vibrational, $\nu$, and rotational, $J$, quantum numbers and depending only on the internuclear distance, $R$. For charged diatomic molecules, the 'charge-modified reduced mass' \cite{Watson1980}, $\mu$ in Eq.~(\ref{vibr}), is defined by
\begin{eqnarray}
    \mu=\frac{m_{1}m_{2}}{m_{1}+m_{2}-m_{e}q}.
\end{eqnarray}
Here, $m_1$ and $m_2$ are the atomic masses, $m_e$ is the mass of an electron, and $q$ is the net charge of the molecule. For simplicity, we further consider only the vibrational spectrum of the ground molecular term, when $J=0$ in Eq.~(\ref{vibr}). Note, that Eq.~(\ref{vibr})  represents a specific scenario where the magnetic field is aligned parallel to the molecular axis and non-adiabatic couplings are omitted. A more general formulation, derived in \cite{Schmelcher1988} (see Eq.~(5) therein), explicitly accounts for both the magnetic field strength and non-adiabatic effects. These terms disappear under the present approximations, resulting in an indirect dependence on B solely through the potential $V(R)$.

To find the vibrational energy levels of a molecule, the second-order differential equation (\ref{vibr}) is solved numerically using a grid of radial points (mesh), as shown in \cite{Stol1, Stol2}. In our implementation, the potential curve $V(R, \theta = 0)$ is interpolated to intermediate grid points using cubic splines. The finite-difference method is then used to perform the integration. To validate our results, we compare the numerically computed eigenvalues and eigenvectors of Eq.~(\ref{vibr}) with those obtained using the \texttt{DUO} package \cite{Yurchenko_2016}, a specialized software intended for calculating the vibrational structure of diatomic molecules. The comparison yielded identical results to within two decimal digits, confirming the accuracy of our implementation.

The results of numerical calculations for the first five rotational energy levels of the X$^1\Sigma^{+}_{g}$ state of H$_2$ and the X$^1\Sigma^{+}$ states of HeH$^+$ and LiH molecules are presented in Table~\ref{tab:vibr_field}. In the absence of a field, the numerical calculations agree well with the available in the literature experimental data and other theoretical calculations. Given that high precision is not required, the values listed in Table~\ref{tab:vibr_field} are rounded to one decimal point.
\begin{table}[h]
    \centering
    \caption{The first five rotational energy levels of the X$^1\Sigma^{+}_{g}$ state of H$_2$ molecule and the X$^1\Sigma^{+}$ states of HeH$^+$, LiH molecules (in cm$^{-1}$) for various field strengths $B^{\parallel}_{0}$ (in atomic units). The second row in each cell of the second column corresponds to the experimental or theoretical values sampled from the literature for comparison. 
   }
    \begin{tabular}{c l l l l l }
    \hline
    & & & H$_{2}$ & &  \\
    \hline
       $B^{\parallel}_{0}$, a.u. &   0.0 &  0.1 & 0.2 & 0.5 & 1.0 \\
       \hline
        $\nu = 0 $    &  0      & 0       & 0       & 0       & 0         \\
        $\nu = 1 $    & 4164.9  & 4190.0  &  4261.5 & 4650.2  & 5457.6    \\
                      & 4148.2$^a$  &   &         &         &             \\
        $\nu = 2 $    & 8101.0  & 8150.5  &  8291.7 & 9054.5 & 10600.3   \\
                      & 8043.4$^{a}$   &   &         &         &         \\
        $\nu = 3 $    & 11805.1 & 11877.7 & 12084.9 & 13199.7 & 15448.0  \\
                      & 11685.6$^{a}$  &   &         &         &         \\
        $\nu = 4 $    & 15274.7 & 15369.5 &  15639.9 & 17094.2 & 20024.9 \\
                      & 15074.9$^{a}$  &   &         &         &         \\     
        $\nu = 5 $    & 18513.3 & 18629.6 & 18961.0 & 20744.8 &  24332.0 \\
                      & 18211.2$^{a}$  &   &         &         &         \\     
     \hline
     & & & HeH$^{+}$ & &  \\
    \hline
    $\nu = 0 $        &  0      & 0       & 0       & 0       & 0       \\
    $\nu = 1 $        & 2905.4  &  2913.6 & 2937.5  & 3092.3  & 3536.3  \\
                      & 2910.9$^{b}$  &   &         &         &         \\
    $\nu = 2 $        & 5512.8  &  5529.1 & 5577.4  & 5887.8 & 6768.5  \\
                      & 5515.1$^{b}$  &   &         &         &         \\
    $\nu = 3 $        & 7824.7  &  7849.2 & 7921.7  & 8385.6  & 9676.1  \\
                      & 7810.7$^{b}$   &   &         &         &        \\
    $\nu = 4 $        & 9829.8  &  9862.4 & 9958.8 &  10573.8 & 12257.1 \\
                      & 9792.8$^{c}$ &   &         &         &          \\
    $\nu = 5 $        & 11518.8 & 11559.6 & 11680.1 & 12446.9 & 14514.7 \\
                      & 11453.2$^{c}$ &   &         &         &         \\
    \hline
     & & & LiH & &  \\
    \hline
    $\nu = 0 $        &  0      & 0        & 0       & 0      & 0       \\ 
    $\nu = 1 $        & 1364.9  &  1392.0  & 1446.8  & 1637.1 & 1953.2  \\
                      & 1359.7$^{d}$&          &         &        &     \\
    $\nu = 2 $        & 2683.8  &  2738.5 & 2848.7  & 3224.3&  3829.2 \\
                      & 2674.6$^{d}$  &          &         &        &   \\
    $\nu = 3 $        & 3958.6  &  4041.7  & 4207.8  & 4764.9 & 5632.2  \\
                      & 3945.5$^{d}$  &          &         &        &   \\
    $\nu = 4 $        & 5190.6  &  5303.0  &  5525.8  & 6261.6 &7367.9 \\
                      & 5173.3$^{d}$  &          &         &        &   \\
    $\nu = 5 $        & 6380.9  &  6523.5  & 6804.0  & 7716.8 & 9042.4\\
                      & 6358.7$^{d}$  &          &         &        &    \\
    \hline
    \end{tabular}
    \label{tab:vibr_field}
    \begin{flushleft}
    \footnotesize{$^a$ values were calculated using potential energy curves fitted using experimentally determined molecular constants \cite{Pingak_2021}};
    \\
    \footnotesize{$^b$ only the first three experimentally determined frequencies are available, the remaining ones are calculated theoretically \cite{Stanke2006}}; 
    \\
    \footnotesize{$^c$ these theoretical values take into account the nonadiabatic correction \cite{Pachucki2012}}; 
    \\
    \footnotesize{$^d$ experimentally defined values \cite{Chan1986}.}
    \end{flushleft}
\end{table}

All values presented in Table~\ref{tab:vibr_field} are calculated as energies with respect to $\nu=0$. This is achieved by generally shifting the PEC so that the energy of the zero vibrational level is zero. According to the results collected in the second column of Table~\ref{tab:vibr_field}, for the lower vibration levels at zero magnetic field our calculation turns out to be quite accurate even without the Born-Oppenheimer correction. For the first two vibration levels, the relative deviation with the results of other works does not exceed one percent. As the vibrational quantum number $\nu$ increases, the accuracy decreases, mainly due to the limitations of the basis sets used. 

From Table~\ref{tab:vibr_field} it can be seen that the vibrational spacings increase systematically with increasing magnetic field, following an almost linear trend at low field strengths and showing a slight upward curvature as the field strength increases. This steady increase in vibrational frequencies reflects the field-induced deepening of the molecular potential well, which strengthens the effective binding and increases the separation between adjacent energy levels.

We further analyzed the effect of a strong magnetic field on the LiH compound and determined the equilibrium distances, $r_{\mathrm{eq}}$, for various field magnitudes with parallel and perpendicular orientations with respect to the molecular axis. The corresponding values  for X$^1\Sigma^{+}$ and a$^3\Sigma^{+}$ states are given in Table~\ref{tab:LiH:req}. 
\begin{table}[h]
      \caption{Comparison of equilibrium distances of the LiH molecule for X$^1\Sigma^{+}$ and  a$^3\Sigma^{+}$ states at various magnetic field strengths. The magnitudes of the field strengths are provided in the first row in atomic units. The values are given for the parallel and perpendicular configurations, denoted as $ r_{\mathrm{eq}}^{\parallel} $ and $ r_{\mathrm{eq}}^{\perp} $, respectively. The signs "$-$" indicate the absence of an equilibrium distance for particular field and state.}
    \begin{tabular}{ c   c   c   c   c   c}
      \hline
 $B_0$, a.u. & 0.1 & 0.2  & 0.5 & 1.0  & 2.0 \\
     \hline  
               &  &    &  a$^3\Sigma^{+}$   &   & \\
\hline               
$r_{\mathrm{eq}}^{\parallel}$ &  $ - $ & $ - $ & $ 1.671 $ & $1.502$ & $0.984$ \\
$r_{\mathrm{eq}}^{\perp}$ &  $ - $ & $ - $ & $ - $ & $1.607$ & $0.997$ \\
\hline  
               &  &   &  X$^1\Sigma^{+}$   &   & \\
     \hline 
$r_{\mathrm{eq}}^{\parallel}$ & $1.586 $ & $ 1.565   $ & $ 1.484$ & $ 1.363 $ & $ 1.199 $ \\
$r_{\mathrm{eq}}^{\perp}$     & $1.572$ & $1.419$ & $1.312$ & $1.063$ &  $0.848$\\    
\hline
      \end{tabular}   
    \label{tab:LiH:req}
\end{table}
These results underscore the significance of magnetic field orientation in modulating molecular binding interactions and highlight the need for further exploration of these phenomena in more complex systems. In particular, our results indicate that in the presence of strong external magnetic fields near astrophysical objects, new bound states can form that are not observed under laboratory conditions.

\section{Discussion and conclusions}
\label{theend}

In this study we employed the coupled cluster method to calculate the electronic structure and properties of light diatomic molecules in strong magnetic fields using modern quantum-chemical packages. To validate our approach, we compared the results obtained with the combined code based on the \texttt{GQCP} and \texttt{PySCF} libraries with outcomes from the \texttt{BAGEL} and \texttt{ChronusQ} packages. Table~\ref{tab:CCvsCI} presents a comparison of the total CCSD energy calculations at equilibrium geometries for the ground state of H$_2$ and HeH$^+$ molecules with those performed within the Full CI approach in \texttt{BAGEL}. To find these energies, the potential energy curves of the ground molecular term at different magnetic field strengths were constructed. We also performed calculations for the first excited singlet and triplet potential energy surfaces. The behavior of the molecular terms as a function of the magnetic field magnitude is shown in Fig.~\ref{fig:4}, \ref{fig:9} and \ref{fig:13} for the compounds H$_2$ HeH$^+$ and LiH, respectively. In addition, we generated a contour plot of the electron charge density for the H$_2$ molecule in its ground state, X$^1\Sigma^+_g$, at various field strengths, see Fig.~\ref{fig:7}. 

The equilibrium bond distances corresponding to the ground molecular term for the H$_2$ HeH$^+$ and LiH compounds were calculated for both zero field conditions and for several strengths related to strong fields, see Table~\ref{tab:req}. The validity of the results was verified by comparing the results obtained with and without the presence of the field, within the framework of different calculation methods and with data available in the literature. 

The minor discrepancies observed in the total energy values presented in Table~\ref{tab:CCvsCI} (on the order of 10$^{-3}$ to 10$^{-4}$ a.u.) can be primarily attributed to the use of the density fitting approximation, also referred to as the resolution-of-identity (RI) method, employed by \texttt{BAGEL} for the calculation of interelectron integrals. The implementation of this approximation is driven by the objective of reducing the computational costs associated with large molecular systems. However, it may introduce numerical inaccuracies of a minor nature. Along with that, the relativistic effects incorporated in the RelFCI treatment within \texttt{BAGEL} are considered negligible at the current level of experimental precision for the systems under investigation.

In addition, the permanent dipole moments of the heteronuclear HeH$^+$ and LiH compounds were calculated. The results are shown in Figs.~\ref{fig:11} and \ref{fig:15}, respectively. To verify our calculations of the PDMs obtained by the combined \texttt{GQCP}+\texttt{PySCF} approach, we compared them with those calculated using the \texttt{ChronusQ} package at the SCF level. As shown in Table~\ref{tab:pdm_comparison}, excellent agreement is achieved within the self-consistent field method. However, the third column in Table~\ref{tab:pdm_comparison} clearly demonstrates the need to consider GIAOs in this type of calculation.  It should be noted that correlation methods for GIAOs are also available in \texttt{ChronusQ} at DFT (density functional theory) and configuration interaction methods, see \cite{Sun2024, Tang2024}. Conversely, the current version of \texttt{BAGEL} does not support density matrices in calculations involving GIAOs, which are crucial for calculating PDMs. Analogous to the calculation of the PDMs, the dependence of the transition dipole moment between the ground and first excited singlet states on the internuclear distance in the H$_2$ and HeH$^+$ molecules was also determined, see Figs.~\ref{fig:6} and \ref{fig:10}, respectively.

This study has led to several important conclusions, particularly on the potential curves and properties of light diatomic molecules. First, as expected, the most energetically preferred state of the molecule relates to the parallel field orientation, with zero angle between the magnetic field direction and the molecular axis, see Fig.~\ref{fig:1}. Second, in the calculation of molecular states, and in particular the ground state, it turns out to be principled to use the gauge origin including atomic orbitals. This is especially true in strong fields where the use of perturbation theory is inappropriate. The method of coupled clusters adapted to GIAOs instead of typical Gaussian orbitals leads to the correct dissociation energy for perpendicular field orientation, see Fig.~\ref{fig:3}. For parallel field orientation, the corresponding phase factors of the wave functions are zero and the dissociation limit can be reached using net Gaussians and the ordinary CC approach. Finally, the PECs, PDMs and TDMs at finite magnetic fields obtained in this work are crucial for calculating cross sections of various reactions and for understanding the formation of molecules near astrophysical objects with strong magnetic fields. Furthermore, this knowledge allows the identification of spectra in a manner similar to the field-free case \cite{Zygelman1989, Zygelman1998}. 

Within the scope of this investigation, we also performed an independent verification of the perpendicular paramagnetic binding mechanism by using the triplet states of hydrogen and lithium hydride molecules as a model system, corroborating the findings reported in \cite{Lange2012}. Our rigorous analysis has also demonstrated that the triplet state of LiH exhibits binding even under a parallel magnetic field orientation, thereby extending the current understanding of field-dependent molecular interactions. In particular, our results indicate that for the LiH molecule, the parallel field orientation can be energetically more favorable for the excited triplet state; see Fig.~\ref{fig:14}, and leads to a robust chemical binding with increasing field. 

It should be noted that all calculations in the present work have been performed in the framework of the non-relativistic approach and the Born-Oppenheimer approximation. The latter is a generally accepted approach for the separate consideration of electronic and nuclear degrees of freedom. However, for light molecules, it is well established that going beyond the adiabatic approximation can be particularly important in precision spectroscopy applications \cite{Handy1986}. In the presence of strong magnetic fields, the Born-Oppenheimer approximation faces additional difficulties because of the emergence of significant non-adiabatic couplings depending on the magnetic field and the lack of proper screening of nuclear charges by the electronic cloud. The influence of non-adiabaticity increases with the field strength and leads to more pronounced deviations, especially as equilibrium bond distances decrease. To accurately describe molecular dynamics and spectroscopic properties under such conditions, it is crucial to incorporate diagonal and, in some cases, off-diagonal non-adiabatic couplings into the theoretical framework \cite{Schmelcher1988}.

In the field-free case, the correction to the electronic energy in the leading order can be accounted for by the Born-Oppenheimer diagonal correction. The latter can be efficiently accounted for in the \texttt{CFOUR} package \cite{cfour}, but only for the zero-field case. For finite magnetic fields, such calculations have only recently been performed, as reported in \cite{Culpitt2022, Monzel2022}. In contrast, in the absence of an external field, our calculated vibrational energy levels exhibit excellent agreement with both theoretical and experimental results from the literature, as detailed in Table~\ref{tab:vibr_field}. Notably, for the lowest two to three vibrational levels in the studied molecular systems, the relative deviation in vibrational energy remains below a fraction of a percent.

In conclusion, this study offers insights into the electronic structure and properties of light diatomic molecules in strong magnetic fields and establishes a basis for further research into the behavior of molecules in extreme environments. Finally, the method developed in the present work for calculating complex electronic structures in finite magnetic fields, employing the \texttt{PySCF} and \texttt{GQCP} libraries, provides a useful and scalable alternative to existing similar codes. The numerical data obtained through this approach can offer significant insights for the interpretation of astronomical observations and elucidate the physical processes that govern the formation and evolution of molecules under such intricate conditions.

\bigskip

\section*{Acknowledgements}
The work of T.Z. was supported by the Foundation for the Advancement of Theoretical Physics and Mathematics "BASIS" (grant No.~23-1-3-31-1). The authors would like to express their gratitude to prof. A. N. Petrov (Petersburg Nuclear Physics Institute) for his insightful discussions and expert advices.

\bibliographystyle{apsrev4-2}
\bibliography{sample}

\end{document}